\documentclass{jfm}

\usepackage{graphicx}
\usepackage{natbib}
\usepackage{float}

\ifCUPmtlplainloaded \else
  \checkfont{eurm10}
  \iffontfound
    \IfFileExists{upmath.sty}
      {\typeout{^^JFound AMS Euler Roman fonts on the system,
                   using the 'upmath' package.^^J}%
       \usepackage{upmath}}
      {\typeout{^^JFound AMS Euler Roman fonts on the system, but you
                   dont seem to have the}%
       \typeout{'upmath' package installed. JFM.cls can take advantage
                 of these fonts,^^Jif you use 'upmath' package.^^J}%
      }
  \else
  \fi
\fi

\ifCUPmtlplainloaded \else
  \checkfont{msam10}
  \iffontfound
    \IfFileExists{amssymb.sty}
      {\typeout{^^JFound AMS Symbol fonts on the system, using the
                'amssymb' package.^^J}%
       \usepackage{amssymb}%
         
         \let\geq=\geqslant
      }{}
  \fi
\fi
\ifCUPmtlplainloaded \else
  \IfFileExists{amsbsy.sty}
    {\typeout{^^JFound the 'amsbsy' package on the system, using it.^^J}%
     \usepackage{amsbsy}}
    {}
\fi

\newsavebox{\astrutbox}
\sbox{\astrutbox}{\rule[-5pt]{0pt}{20pt}}

\title{Viscous-poroelastic interaction as mechanism to create adhesion in frogs' toe pads}
\shorttitle{Viscous-elastic adhesion}
\author[A. Tulchinsky and A. D. Gat]{A. Tulchinsky and A. D. Gat}
\affiliation{Faculty of Mechanical Engineering, Technion - Israel Institute of Technology, Haifa 32000, Israel}

\pubyear{2014} \volume{999} \pagerange{999--9999}
\date{2014}
\begin{document}

\maketitle

\begin{abstract}
The toe pads of frogs consist of soft hexagonal structures and a viscous liquid contained between and within the hexagonal structures. It has been hypothesized that this configuration creates adhesion by allowing for long range capillary forces, or alternatively, by allowing for exit of the liquid and thus improving contact of the toe pad. In this work we suggest interaction between viscosity and elasticity as a mechanism to create temporary adhesion, even in the absence of capillary effects or van der Waals forces. We initially illustrate this concept experimentally by a simplified configuration consisting of two surfaces connected by a liquid bridge and elastic springs. We then utilize poroelastic mixture theory and model frog's toe pads as an elastic porous medium, immersed within a viscous liquid and pressed against a rigid rough surface. The flow between the surface and the toe pad is modeled by the lubrication approximation. Inertia is neglected and analysis of the elastic-viscous dynamics yields a governing partial differential equation describing the flow and stress within the porous medium. Several solutions of the governing equation are presented and show a temporary adhesion due to stress created at the contact surface between the solids. This work thus may explain how some frogs (such as the torrent frog) maintain adhesion underwater and the reason for the periodic repositioning of frogs' toe pads during adhesion to surfaces.
\end{abstract}

\section{Introduction}
The toe pads of frogs consist of soft thin hexagonal structures and a viscous fluid between and within the soft structures \citep{ernst1973digitalI,ernst1973digitalII,green1979treefrog}. It has been hypothesized that such configuration enables attachment to surfaces by capillary forces \citep{Emerson1980Toe,HANNA1991Adhesion,Federle2006Why}, or alternatively that the channel network allows for exit of the viscous liquid and thus improves contact of the toe pad with the surface \citep{Federle2006Wet,persson2007wet,tsipenyuk2014use}. Analytical works include \cite{Federle2006Wet} who modeled the contribution of capillary forces to shear stress on the surface and \cite{persson2007wet}, who studied the effect of elasticity on capillary forces.

In this work we suggest a mechanism for creating temporary adhesion based on interaction between viscous flow and elastic deformation. We model the toe pads of frogs as an elastic porous medium \citep[similarly to][]{battiato2010elastic,battiato2012self}, immersed within a viscous liquid, and pressed against a solid surface with known roughness. The dynamics of the elastic porous material are studied via the poroelastic theory \citep{Biot1972Theory,Bowen1980Incompressible,Ambrosi2000Modeling}. The flow between the frogs' toe pads and the solid surface is modeled by the lubrication approximation. Forces and kinematic constraints acting on the liquid-saturated toe pad deform the material. The deformation of the porous material creates a viscous flow within the toe pads while modifying its stress field. The viscous fluid, flowing from the lubrication region into the porous material, yields a pressure-field and thus effectively creates a force acting between the porous material and the surface, perpendicular to the surface. This force creates tangential friction between the porous material and the solid surface, thus preventing slip on the surface. Such a mechanism will allow for adhesion even in the absence of capillary forces and thus may explain how some frogs can maintain adhesion in the presence of rain or while being submerged \citep[such as river frogs, see][]{endlein2013stickinglike,barnes2002bionics}. In addition, the time-varying nature of the suggested mechanism may explain why frogs periodically reposition their toes when connected to a surface \citep{endlein2013stickingunder,endlein2013stickinglike}.

The structure of this paper is as follows: In the next section we illustrate the concept experimentally by a simplified configuration consisting of two surfaces connected by a liquid bridge and elastic springs. In section 3.1 we define the poroelastic problem. In section 3.2 we analyze the flow-field and deformation field within the poroelastic medium. In section 3.3 we analyze the flow in the lubrication region between the toe pad and the solid surface and in section 3.4 we obtain a governing equation for the dynamics of the toe pad, present several solutions and estimate viscous-poroelastic adhesion in frogs. In section 4 we summarize the results.

\section{Experimental illustration of viscous-elastic friction creation}
\begin{figure}
\centering
\includegraphics[width=1\textwidth]{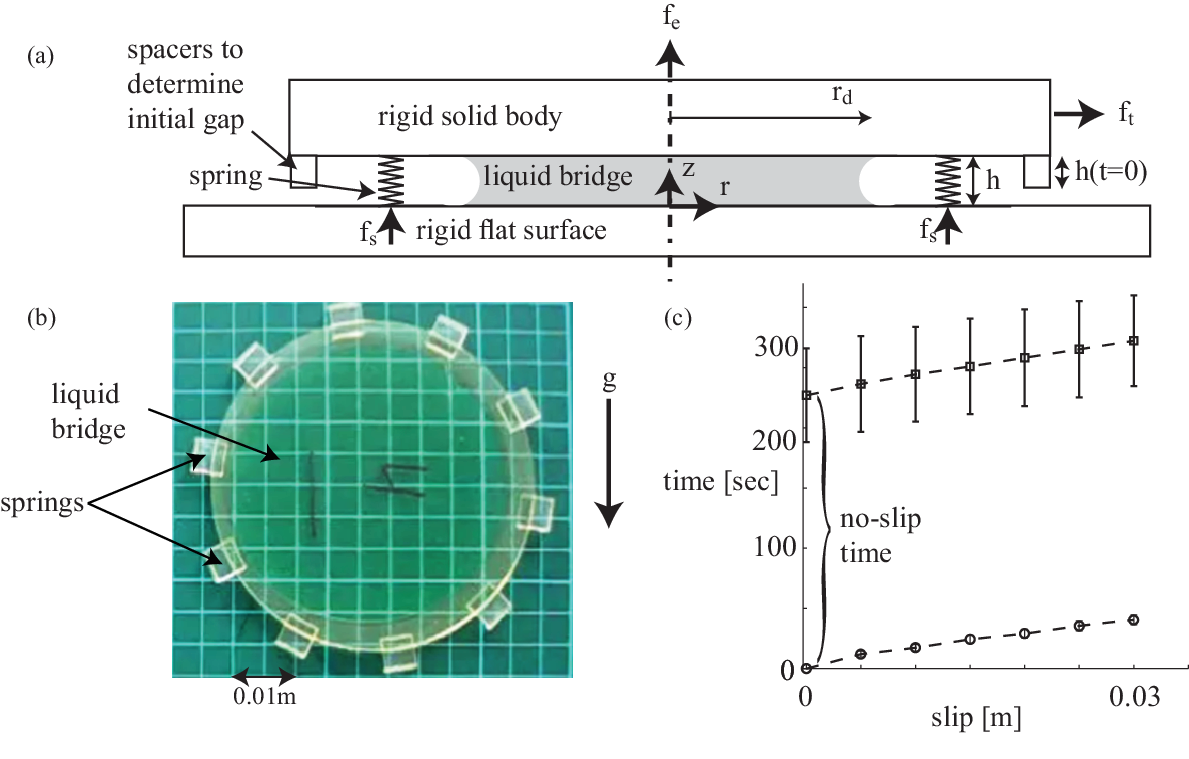}
\caption{(a) An illustrative front view of the mathematical model. 
(b) An upper view photo of the experimental model attached to a horizontal surface with a liquid bridge between and (c) experimental results presenting time vs. slip for plates with springs (squares) and without springs (circles). Bars presents standard deviation.}
\label{Figure_1}
\end{figure}

We initially focus on a simplified case in order to illustrate the effects of viscous-elastic interaction on friction. We examine the dynamics of two parallel surfaces connected by a liquid bridge and linearly elastic springs (see Fig. \ref{Figure_1}a). While in this case there are no poroelastic dynamics, it includes both viscous forces and elastic forces due to the liquid bridge and the linear springs, respectively. The springs are located outside of the liquid bridge and do not affect the flow-field. Friction is created by the normal force applied by the springs on the contact area with the surface. The configuration is axi-symmetric.
We denote the gap between the two surfaces $h$, the viscosity $\mu$, the surface tension $\gamma$, the density $\rho$, the liquid drop volume $v_d$, the liquid radius $r_d$, the total springs 
stiffness $k$, the relaxed spring length $h_r$ (thus the normal force applied by the springs is $k(h-h_r)$) and the external normal force $f_e$ (see Fig. \ref{Figure_1}a). Hereafter, characteristic 
values are denoted by asterisk superscripts. We define $u^*$ as the characteristic speed, $p^*$ as the characteristic pressure and $h^*$ as the characteristic gap. Following \cite{Gat.2011}, under the assumptions of shallow liquid bridge, $h^*/r_d\ll 1$, negligible inertia of the liquid, $h^{*2}\rho u^*/\mu \ll 1$, negligible gravity $g\rho h^*/p^*\ll 1$ and negligible capillary forces $\gamma v_d/(h^*)^2\ll1$, the force balance equation is 

\begin{equation}
f_e-\mu \frac{3v_d^2}{2\pi h^5}\frac{dh}{dt}+k(h_r-h)=0.
\end{equation}
The tangential friction acting at the solid interface is $f_t=\mu_s(h_r-h)k$, where $\mu_s$ is the friction coefficient. For cases in which both the springs and viscosity terms are non-negligible, order of magnitude yields the characteristic time scale $t^*=\mu v_d^2/\pi^2 k h^{*5}$. During this time scale the springs will apply a normal force on the surface of the order of $\approx k h^*$, which will in turn create a tangential friction force. 

We conducted experiments with a circular flat plate connected to $10$ extended beams acting as linear elastic springs (see Fig. \ref{Figure_1}b). The solid was printed by Objet Eden250, and the material is Objet FullCure720. In all cases the configuration was tested immediately after printing and used only once. The liquid is silicon oil droplet with volume $v_d=5\cdot10^{-6}m^3$, density $\rho=750kg/m^3$, surface tension $0.021N m^{-1}$ and viscosity $\mu=60Pa s$. The plate was pressed against a rigid surface (positioned parallel to gravity, see Fig. \ref{Figure_1}b) and released at $t=0$. The properties of the examined configuration are $k\approx 4.5\cdot 10^4 N/m$, $h_r=1.3mm$. The order of magnitude of capillary forces is $2\gamma v_d/(h^*)^2\approx 0.2N$ and is negligible compared with the order of magnitude of the elastic force $kh^*\approx 50N$ (where $h^*=1mm$). For such a configuration the characteristic time scale of is $t^*=\mu v_d^2/\pi^2 k h^{*5}=O(10^3s)$. We examined $6$ identical plates and $6$ control plates without springs. In all cases the initial gap was determined by spacers (see Fig. \ref{Figure_1}a) as $h(t=0)=0.8mm$. The results are presented in Fig \ref{Figure_1}c and clearly show enhancement of friction due to viscous-elastic interaction. While the control plates without springs immediately slipped, the configurations with springs slipped only at $\approx 250s$. 

\section{Adhesion due to viscous-poroelastic interaction}
We now turn to analyze a more complex case of viscous-elastic interaction involving viscous flow within an elastic porous material as a mechanism to create adhesion. 

\subsection{Problem Definition}
\begin{figure}
\includegraphics[width=0.9\textwidth]{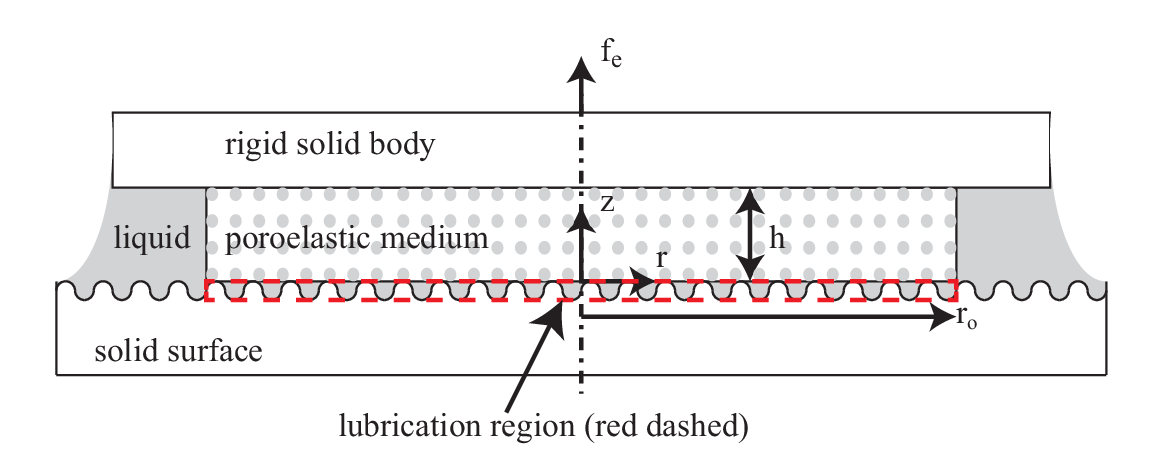}
\centering
\caption{Illustrative description of the model consisting of a poroelastic material (dotted area) connecting a rigid body and a lubrication region (red dashed). The poroelastic material is axi-symmeteric with radius $r_o$, height $h$ and external axial force $f_e$ acting on the rigid body.}
\label{figure_2}
\end{figure}

We model the interaction between an elastic porous material and a Newtonian fluid via the poroelastic mixture theory \citep{Preziosi1996Infiltration,Bowen1980Incompressible,atkin1976continuum,rajagopalmechanics}. We focus on axi-symmetric configurations with negligible inertial and capillary effects. The relevant regions are the poroelastic region and a thin liquid-filled gap between the poroelastic region and the rough surface (marked by red dotted line, see Fig. \ref{figure_2}). The relevant variables are the radius of the poroelastic material $r_o$, the poroelastic material height $h$, the poroelastic material relaxed height $h_r$ (height when no stress is applied on the solid), the external force $f_e$, the solid fraction (i.e. ratio of solid volume to total volume of liquid and solid) $\Phi$, the relaxed solid fraction $\Phi_r$, the time $t$, the radial and axial coordinates $\mathbf{r}=(r,z)$, the solid velocity $\mathbf{v_s}=(u_s,v_s,w_s)$, the liquid velocity $\mathbf{v_l}=(u_l,v_l,w_l)$, the liquid viscosity  $\mu$, the solid density $\rho_s$, the liquid density $\rho_l$, the permeability $\mathbf{k}$, and the excess stress tensor $\mathbf{\sigma}$ \citep[defined as the stress tensor of the mixture minus the liquid pressure, see][]{Preziosi1996Infiltration}. 

\subsection{Analysis of the poroelastic medium}
The dynamics of a poroelastic medium saturated with Newtonian incompressible liquid is governed by the conservation of mass of the solid 
\begin{equation}
\frac{\partial\Phi}{\partial t}+\nabla\cdot(\Phi\mathbf{v_s})=0,
\label{eq1}
\end{equation}
the conservation of mass of the liquid
\begin{equation}
-\frac{\partial\Phi}{\partial t}+\nabla\cdot[(1-\Phi)\mathbf{v_l}]=0,
\label{eq2}
\end{equation}
Darcy's law for the flow within the porous material
\begin{equation}
\mathbf{v_l}-\mathbf{v_s}=-\frac{\mathbf{k}}{(1-\Phi)\mu}(\nabla p-\rho_l\mathbf{g})
\label{eq3}
\end{equation}
and equation of the mixed momentum
\begin{equation}
\rho_c\left(\frac{\partial\mathbf{v_m}}{\partial t}+\mathbf{v_m}\cdot\nabla\mathbf{v_m}\right)=-\nabla p+\nabla\cdot\mathbf{\sigma}+[\rho_s\Phi+\rho_l(1-\Phi)]\mathbf{g}
\label{eq4}
\end{equation}
where $\rho_c=\Phi\rho_s+(1-\Phi)\rho_l$ is the density of the mixture considered as a whole and $\mathbf{v_m}=[\Phi\rho_s\mathbf{v_s}+(1-\Phi)\rho_l\mathbf{v_l}]/\rho_c$ is the mass average velocity \citep[see][]{Preziosi1996Infiltration} .

The relevant boundary conditions are no solid velocity at $z=0$ and external velocity function $\mathbf{v_e}(t)$ at $z=h(t)$,
\begin{equation}
\mathbf{v_s}(z=0)=0,\,\,\,\mathbf{v_s}(z=h(t))=\mathbf{v_l}(z=h(t))=\mathbf{v_e}(t).
\label{bc}
\end{equation} 
The initial conditions are 
\begin{equation}
\Phi(t=0)=\Phi_i,\,\,\,h(t=0)=h_i.
\end{equation} 
 
Hereafter all normalized parameters are denoted by capital letters and characteristic values are denoted by asterisk superscripts. We define normalized coordinates $R,Z$ and time $T$
\begin{equation}\label{nor1}
R=\frac{r}{r_o},\,\,\,Z=\frac{z}{h^*},\,\,\,T=\frac{t}{h^*/w^*}, 
\end{equation}
normalized liquid velocity $\mathbf{V_l}$ and pressure $P$
\begin{equation}\label{nor2}
\mathbf{V_l}=\frac{\mathbf{v_l}}{w^*},\,\,\, P=\frac{p}{p^*},
\end{equation}
normalized solid velocity $\mathbf{V_s}$, solid stress $\mathbf{\Sigma}$, permeability $\mathbf{K}$
\begin{equation}
\mathbf{V_s}=\frac{\mathbf{v_s}}{w^*},\,\,\, \mathbf{\Sigma}=\frac{\mathbf{\sigma}}{\sigma^*},\,\,\, 
\mathbf{K}=\frac{\mathbf{k}}{k^*},
\end{equation}
and poroelastic height $H$
\begin{equation}
H=\frac{h(t)}{h^*},
\end{equation}
where $h^*$ is the characteristic height, $w^*$ is the characteristic speed,  $p^*$ is the characteristic pressure, $\sigma^*$ is the characteristic axial stress and $k^*$ is the characteristic permeability. 

Our analysis will focus on a thin geometry, defined as
\begin{equation}
\varepsilon_1=\frac{h^*}{r_o}\ll1.
\end{equation}
Substituting the normalized variables yield the leading order form of (\ref{eq1}-\ref{eq4}) as
\begin{equation}\label{eq:liqmascon}
\frac{\partial\Phi}{\partial T}+\frac{\partial}{\partial Z}(\Phi W_s)+O(\varepsilon_1)=0,
\end{equation}
\begin{equation}\label{eq:solmascon}
-\frac{\partial\Phi}{\partial T}+\frac{\partial}{\partial Z}[(1-\Phi) W_l]+O(\varepsilon_1)=0,
\end{equation}
\begin{equation}\label{eq:dar}
W_l-W_s=-\frac{k^*p^*}{\mu w^*h^*}\frac{K_{zz}}{(1-\Phi)}\frac{\partial P}{\partial Z}+O\left(\varepsilon_1,\frac{\rho_l g k_0}{\mu w^*}\right)
\end{equation}
and
\begin{equation}\label{eq:mom}
\frac{\partial P}{\partial Z}=\frac{\sigma^*}{p^*}\frac{\partial\Sigma_{zz}}{\partial Z}+O\left(\varepsilon_1,\frac{\rho_j (w^*)^2}{p^*}, \frac{[\rho_s\Phi+\rho_l(1-\Phi)]g}{p^*}\right),
\end{equation} 
where $\Sigma_{zz}$ and $K_{zz}$ are the axial terms of $\mathbf{\Sigma}$ and $\mathbf{K}$, respectively, and $W_s$ and $W_l$ are the axial solid and liquid speeds, respectively.

The constitutive relationships for the permeability, $k$, and stress $\sigma$ are approximated as \citep[following previous works of][]{Anderson2005Imbibition,siddique2009newtonian,Siddique2009Capillary}
\begin{equation}\label{eq:streconst}
k_{zz}=\frac{k_0}{\Phi},\,\,\, \sigma=m(\Phi_r-\Phi)
\end{equation}
where $k_0>0$ and $m>0$ are known constants defining the poroelastic material permeability and stiffness, respectively. Order of magnitude of (\ref{eq:mom}) and (\ref{eq:streconst}) yields $p^*=\sigma^*=m$ and $k^*=k_0$. 

We define the scaled coordinate $\Psi=z/h(t)$ and formulate the problem according to \cite{Preziosi1996Infiltration}. Adding (\ref{eq:solmascon}), (\ref{eq:liqmascon}), integrating over $Z$ and using (\ref{bc}) we obtain
\begin{equation}\label{eq:mixcon}
W_l(1-\Phi)+\Phi W_s=\frac{\partial H}{\partial T}.
\end{equation}
From (\ref{eq:dar}) and (\ref{eq:mixcon}),
\begin{equation}\label{eq:ws}
W_s=\frac{\partial H}{\partial T}-\frac{mk_0}{\mu w^*h^*}\frac{1}{\Phi H(T)}\frac{\partial \Phi}{\partial \Psi}
\end{equation} 
and
\begin{equation} \label{eq:wl}
W_l=\frac{\partial H}{\partial T}+\frac{mk_0}{\mu w^*h^*}\frac{1}{(1-\Phi)H(T)}\frac{\partial \Phi}{\partial \Psi}.
\end{equation} 
Combining (\ref{eq:wl}) and (\ref{eq:liqmascon}) together with (\ref{eq:mom}) yields
\begin{equation}\label{eq:PDE}
\frac{\partial \Phi}{\partial T}+(1-\Psi)\frac{\partial H}{\partial T}\frac{1}{H(T)}\frac{\partial \Phi}{\partial \Psi}=\frac{mk_0}{\mu w^*h^*}\frac{1}{H^2(T)}\frac{\partial^2 \Phi}{\partial \Psi^2}.
\end{equation} 
We express the boundary conditions (\ref{bc}) in terms of solid fraction $\Phi$ as
\begin{equation}
\frac{mk_0}{\mu w^* h^*}\frac{\partial \Phi}{\partial \Psi}=0,\,\,\, \Psi=1
\end{equation}
and
\begin{equation}\label{bc2}
\frac{mk_0}{\mu w^* h^*}\frac{\partial \Phi}{\partial \Psi}-H(T)\frac{\partial H}{\partial T}\Phi=0,\,\,\, \Psi=0,
\end{equation}
where $\mu w^* h^*/mk_0$ can be interpreted as the ratio between the characteristic speed of the poroelastic problem, $w^*$, and the characteristic speed of the viscous flow, $mk_0/\mu h^*$.

\subsection{Analysis of lubrication region}
We denote by tildes the liquid velocity $\mathbf{\tilde{v}_l}=(\tilde{u}_l,\tilde{w}_l)$ in the lubrication region, where $\tilde{u}_l$ is the radial speed and $\tilde{w}_l$ is the axial speed. The governing equations in the lubrication region are the axi-symmetric Stokes equations for Newtonian incompressible liquid
\begin{equation}
\nabla p=\mu\nabla^2 \mathbf{\tilde{v}_l}
\label{moment}
\end{equation} 
and conservation of mass
\begin{equation}
\label{massconserve}
\nabla\cdot\mathbf{\tilde{v}_l}=0.
\end{equation} 
The relevant boundary conditions are gauge pressure $p=0$ at $r=r_o$, symmetry $\partial p/\partial r=0$ at $r=0$, no-slip and no-penetration $\mathbf{\tilde{v}_l}=0$ at the solid interface and mass-flux and pressure matching to the poroelastic region $\tilde{w}_l=(1-\Phi)w_l$ at $z=0$. The liquid slip at the boundary of the porous material, $z=0$, is proportional to $\propto\sqrt{k_0} \partial \tilde{u}_l / \partial z$, where $\sqrt{k_0}$ is the characteristic length scale of the permeable material \citep{Beavers.1967}.  We relate the characteristic roughness $\tilde{h}$ to average viscous resistance by defining 
\begin{equation}\label{hdif}
\tilde{h}^{-3}=\frac{1}{A}\int_A{h_s^{-3} dA},
\end{equation}
 where $h_s$ is the local gap between the solid surface and the poroelastic surface and A is a sufficiently large surface  area on the z plane. The normalized velocity $\tilde{U_l},\tilde{W_l}$ and $\tilde{Z}$ coordinate for the lubrication region are defined as
\begin{equation}
\tilde{U_l}=\frac{u}{u^*},\,\,\, \tilde{W_l}=\frac{\tilde{W_l}}{w^*},\,\,\,\tilde{Z}=\frac{z}{\tilde{h}}.
\label{nor_val}
\end{equation}

We require sufficiently small $k_0$ so that 
\begin{equation}\label{small}
\varepsilon_2=\frac{k_0 h^*}{\tilde{h}^{3}/12}\ll1,\,\,\,\varepsilon_3=\frac{u_l(Z=0)}{u^*}=\frac{\sqrt{k_0}}{h^*}\ll1
\end{equation} 
representing the scaled pore size and the slip at the boundary of the poroelastic surface. For $\varepsilon_3\ll1$ we require $\tilde{U}_l=0$ at $\tilde{Z}=0$. \citep[The lubrication region may also be modeled as a porous region of depth $h_p$ and the radial permeability $k_p$, where $\tilde{h}^{3}/12=k_p h_p$, see][]{battiato2010elastic,battiato2012self}.

Substituting (\ref{nor_val}) into (\ref{moment},\ref{massconserve}), order of magnitude analysis yields
\begin{equation}\label{wstar}
u^*=\frac{w^*r_o}{\tilde{h}},\,\,\,w^*=\frac{m \tilde{h}^{3}/12 }{\mu r_0^2},
\end{equation}
and normalized (\ref{moment}) is thus
\begin{equation}
\label{eq_norm_mom_r}
\frac{\partial P}{\partial R}\sim \frac{1}{H_s^3}\int_{-H_s}^{0}{\tilde{U_l} d\tilde{Z}},\,\,\,\frac{\partial P}{\partial \tilde{Z}}\sim0,
\end{equation}
where $H_s=h_s/\tilde{h}$ is the local normalized gap. We substitute (\ref{eq_norm_mom_r}) into (\ref{massconserve}) and apply the boundary conditions $\tilde{W_l}=(1-\Phi)W_l$ at $\tilde{Z}=0$ (flux matching between the lubrication region and the poroelastic region) and $\tilde{U_l}=\tilde{W_l}=0$ at the solid surface. Thus we obtain the pressure distribution in the lubrication region. Substituting into the variables of the poroelastic region and utilizing (\ref{hdif}) yields
\begin{equation}
P(\Psi=0)\sim-(1-\Phi(\Psi=0))W_l(\Psi=0)\frac{1-R^2}{4}.
\label{lub_pre}
\end{equation}

\subsection{Results}
\subsubsection{Governing equations}
Substituting (\ref{wstar}) into (\ref{eq:PDE}-\ref{bc2}) yields 
\begin{equation}
\frac{m k_0}{\mu w^* h^*}=\frac{k_0 h^*}{\tilde{h}^{3}/12}\frac{r_o^2}{(h^*)^2}=\frac{\varepsilon_2}{\varepsilon_1^2},
\end{equation} 
and the governing equation
\begin{equation}\label{governing_final_phi}
\frac{1}{H^2}\frac{\partial^2 \Phi}{\partial \Psi^2}-\frac{\varepsilon_1^2}{\varepsilon_2}\left[\frac{\partial \Phi}{\partial T}+(1-\Psi)\frac{\partial H}{\partial T}\frac{1}{H}\frac{\partial \Phi}{\partial \Psi}\right]=0,
\end{equation} 
with the boundary conditions
\begin{equation}\label{governing_final_phi_bc1}
\frac{\partial \Phi}{\partial \Psi}-\frac{\varepsilon_1^2}{\varepsilon_2}H\frac{\partial H}{\partial T}\Phi=0,\,\,\, \Psi=0,
\end{equation}
\begin{equation}\label{governing_final_phi_bc2}
\frac{\partial \Phi}{\partial \Psi}=0,\,\,\, \Psi=1
\end{equation}
and initial condition
\begin{equation}\label{initialc}
\Phi=\Phi_i,\,\,\, T=0.
\end{equation}
The liquid speed $W_l$ and solid speed $W_s$ are thus
\begin{equation}\label{eq:ws_final}
W_s=\frac{\partial H}{\partial T}-\frac{\varepsilon_2}{\varepsilon_1^2}\frac{1}{\Phi H(T)}\frac{\partial \Phi}{\partial \Psi},\,\,\,W_l=\frac{\partial H}{\partial T}+\frac{\varepsilon_2}{\varepsilon_1^2}\frac{1}{(1-\Phi)H(T)}\frac{\partial \Phi}{\partial \Psi}.
\end{equation} 
Substituting (\ref{eq:ws_final}) into (\ref{lub_pre}) yields the gauge pressure at $Z=0$ 
\begin{equation}
P(Z=0)=-\left(1-\Phi(0)\right)\left(\frac{\partial H}{\partial T}+\frac{\varepsilon_2}{\varepsilon_1^2}\frac{1}{(1-\Phi(0))H}\frac{\partial \Phi(0)}{\partial \Psi}\right)\frac{1-R^2}{4}.
\label{pressure_sol}
\end{equation} 
Integration over $P(Z=0)$ will yield the total normal force applied by the liquid. The external force applied on the poroelastic material $F_e=f_e/mr_o^2$ (see Fig. 2), is balanced by the force applied by the solid surface (denoted hereafter as $F_s=\pi(\Phi_r-\Phi(\Psi=0))$ and normalized by $mr_o^2$) and the force applied by the liquid gauge pressure at $\Psi=0$. The force balance equation, $F_e=\int_0^1{(\Sigma_{zz}-P)}2\pi RdR$, yields the expression for the external force
\begin{equation}
F_e=\pi(\Phi_r-\Phi(\Psi=0))+\frac{\pi}{8}\left(1-\Phi(0)\right)\left(\frac{\partial H}{\partial T}+\frac{\varepsilon_2}{\varepsilon_1^2}\frac{1}{(1-\Phi(0))H}\frac{\partial \Phi(0)}{\partial \Psi}\right).
\label{Dynamics}
\end{equation} 

For the limit $\varepsilon_1^2/\varepsilon_2\ll1$ we define the asymptotic expansion
\begin{equation}\label{phi_expansion}
\Phi=\Phi_0+\left(\frac{\varepsilon_1^2}{\varepsilon_2}\right)\Phi_1+\left(\frac{\varepsilon_1^2}{\varepsilon_2}\right)^2\Phi_2+O\left[\left(\frac{\varepsilon_1^2}{\varepsilon_2}\right)^3\right].
\end{equation}
(We focus on the limit $\varepsilon_1^2/\varepsilon_2\ll1$ since it agrees with the characteristic physical parameters of frogs' toe pads, see Section 3.4.4.) We substitute (\ref{phi_expansion}) into (\ref{governing_final_phi}-\ref{initialc}). The leading order problem yields $\Phi_0(T)$ is a function of time only. The first order terms of (\ref{governing_final_phi}) and (\ref{governing_final_phi_bc2}) yields 
\begin{equation}\label{phi_1}
\Phi_1=H^2 \frac{\partial \Phi_0}{\partial T}\left(\frac{\Psi^2}{2}-\Psi\right)+C_1(T).
\end{equation}
Substituting $\Phi_0$ and $\Phi_1$ into the $O(\varepsilon_1^2/\varepsilon_2)$ order boundary condition at $\Psi=0$ (\ref{governing_final_phi_bc1}) yields an ordinary differential equation with regard to time
\begin{equation}\label{bceq}
\frac{\partial \Phi_0}{\partial T}+\frac{\partial H}{\partial T}\frac{\Phi_0}{H}=0.
\end{equation} 
Solving (\ref{bceq}) with (\ref{initialc}) yields $\Phi_0=\Phi_i H(0)/H(T)$. We determine $C_1(T)$ from the $O[(\varepsilon_1^2/\varepsilon_2)^2]$ order (\ref{governing_final_phi}),
\begin{equation}\label{o2}
\frac{\partial^2 \Phi_2}{\partial \Psi^2}=H^2\frac{\partial \Phi_1}{\partial T}+(1-\Psi)\frac{\partial H}{\partial T}H\frac{\partial \Phi_1}{\partial \Psi} 
\end{equation}
and the $O[(\varepsilon_1^2/\varepsilon_2)^2]$ boundary and initial conditions. Substituting (\ref{phi_1}) into (\ref{o2}) and applying (\ref{governing_final_phi_bc2}) we obtain an ordinary differential equation from the boundary condition at $\Psi=0$, similarly to (\ref{bceq}), for the $O(\varepsilon_1^2/\varepsilon_2)$ order,
\begin{equation}\label{order_2}
\frac{\partial C_1}{\partial T}+\frac{\partial H}{\partial T}\frac{1}{H}C_1=-\frac{1}{3}\frac{\Phi_i H(0)}{H}\frac{\partial}{\partial T}\left(H\frac{\partial H}{\partial T}\right).
\end{equation}
Applying (\ref{initialc}) we solve (\ref{order_2}) and obtain,
\begin{equation}
C_1=-\frac{H(0)}{3}\Phi_i  \frac{\partial H}{\partial T}.
\end{equation}
Thus $\Phi$ to order $O(\varepsilon_1^2/\varepsilon_2)$ is
\begin{equation}
\label{phi_sol}
\frac{\Phi}{\Phi_i}\sim\frac{H(0)}{H}+\frac{\varepsilon_1^2}{\varepsilon_2} H(0)\frac{\partial H}{\partial T}\left(\Psi-\frac{\Psi^2}{2}-\frac{1}{3}\right),
\end{equation}
and from (\ref{initialc}) we obtain the requirement $\partial H(0)/\partial T=0$ in order to satisfy spatially uniform initial condition.

\subsubsection{Externally controlled $H$ for $\varepsilon_1^2/\varepsilon_2\ll1$}
\begin{figure}
\includegraphics[width=0.8\textwidth]{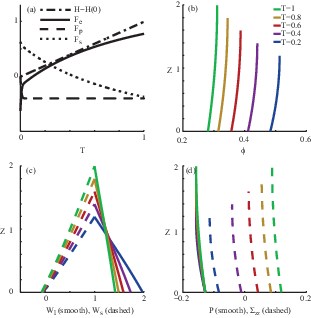}
\centering
\caption{Configuration with speed $\partial H/\partial T=tanh(100T)$ (effectively $\partial H/\partial T\approx1$ for $T>0.02$), initial uniform solid fraction of the poroelastic medium $\Phi_i=0.6$, relaxed solid fraction $\Phi_r=0.4$, and $\varepsilon_1^2/\varepsilon_2=0.1$. Part (a) presents change in gap $H-1$ (dash-dotted), external required force $F_e$ (solid), total force by the liquid pressure $F_p$ (dashed) and force applied by the solid surface $F_s$ (dotted). In parts (b-d) blue, purple, red, yellow and green lines mark normalized time $T=0.2$, $0.4$, $0.6$, $0.8$, and $1$, respectively. Part (b) presents solid fraction $\Phi$ vs. $Z$ for various times. Part (c) presents liquid speed $W_l$ (solid) and solid speed $W_s$ (dashed) vs. $Z$ for various times. Part (d) presents pressure $P$ (solid) and axial stress $\Sigma_{zz}$ vs. $Z$ for various times.}
\label{Figure_3}
\end{figure}

Based on (\ref{phi_sol}) and (\ref{Dynamics}) we can calculate $F_e$ required to achieve an arbitrary $H(T)$. After calculating $H(T)$, exact solutions of the solid fraction $\Phi$, fluid velocity $W_l$, solid velocity $W_s$ and internal stress $\Sigma$ can be obtained from (\ref{phi_sol}, \ref{eq:ws_final}-\ref{Dynamics}). Figure \ref{Figure_3} presents a case of $\partial H/\partial T=tanh(CT)$, where $C=100$ and thus $\partial H/\partial T\approx1$ for $T>0.02$. The initial uniform solid fraction of the poroelastic medium is $\Phi_i=0.6$, relaxed solid fraction is $\Phi_r=0.4$, and $\varepsilon_1^2/\varepsilon_2=0.1$. Part (a) presents change in poroelastic layer height $H-1$ (dash-dotted), external force $F_e$ (solid), total force by the liquid pressure $F_p$ (dashed) and force applied by the solid surface $F_s$ (dotted). The total force by the liquid pressure $F_p$ becomes nearly constant from $T\approx 0.03$ (as the $\partial H/\partial T\rightarrow 1$), while the $F_s$ decreases in time. The external force is initially negative and prevents the compressed poroelastic material to expand more rapidly than the required $\partial H/\partial T=1$. From $T\approx 0.1$ $F_e$ is positive and acts to increase $\partial H/\partial T$. In parts (b-d) blue, purple, red, yellow and green lines mark normalized time $T=0.2$, $0.4$, $0.6$, $0.8$, and $1$, respectively. Part (b) presents solid fraction $\Phi$ vs. $Z$ for various times, showing decrease of $\Phi$ with time and as $Z\rightarrow 0$. Part (c) presents liquid speed $W_l$ (solid) and solid speed $W_s$ (dashed) vs. $Z$ for various times. Both speeds are approximately linear with $Z$ with gradients decreasing with time. Part (d) presents pressure $P$ (solid) and axial stress $\Sigma_{zz}$ vs. $Z$ for various times. The average pressure $P$ is approximately constant, due to the uniform liquid mass-flux into the poroelastic material. The axial stress increases with time due to the forced deformation of the poroelastic material. Due to the initial compression, the axial stress is negative until $T\approx0.4$ and then becomes positive due to stretching by the external force.

\subsubsection{Externally controlled $F_e$ for $\varepsilon_1^2/\varepsilon_2\ll1$}
We apply (\ref{phi_sol}) with (\ref{Dynamics}) to obtain $H$ as a function of the external force acting on poroelastic material for $\varepsilon_1^2/\varepsilon_2\ll1$,
\begin{equation}\label{eq:extst}
F_e=\pi\left(\Phi_r-\Phi_i\frac{H(0)}{H}\right)+\frac{\pi}{8}\frac{\partial H}{\partial T}.
\end{equation}
For constant $F_e$ separation of variables yields an implicit solution of $H(T)$,
\begin{equation}\label{sol_fe}
T=\frac{\pi}{8}\left[\frac{H-H(0)}{F_e-\pi\Phi_r}-\frac{\pi \Phi_i H(0)}{(F_e-\pi\Phi_r)^2}\ln\left(\frac{H(F_e-\pi\Phi_r)-\pi\Phi_i H(0)}{H(0)(F_e-\pi\Phi_r)-\pi\Phi_i H(0)}\right)\right].
\end{equation}

\begin{figure}
\includegraphics[width=0.8\textwidth]{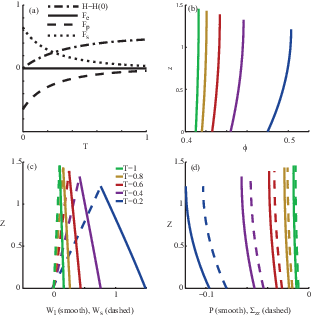}
\centering
\caption{Configuration with initial compression and a sudden release at $T=0$ with $F_e=0$. The initial uniform solid fraction of the poroelastic medium is $\Phi_i=0.6$, relaxed solid fraction $\Phi_r=0.4$, and $\varepsilon_1^2/\varepsilon_2=0.1$. Part (a) presents change in poroelastic layer height $H-1$ (dash-dotted), external required force $F_e$ (solid), total force by the liquid pressure $F_p$ (dashed) and force applied by the solid surface $F_s$ (dotted). In parts (b-d) blue, purple, red, yellow and green lines mark normalized time $T=0.2$, $0.4$, $0.6$, $8$, and $1$, respectively. Part (b) presents solid fraction $\Phi$ vs. $Z$ for various times. Part (c) presents liquid speed $W_l$ (solid) and solid speed $W_s$ (dashed) vs. $Z$ for various times. Part (d) presents pressure $P$ (solid) and stress $\Sigma_{zz}$ vs. $Z$ for various times.}
\label{Figure_4}
\end{figure}

Figure \ref{Figure_4} presents the case of initial compression of the poroelastic material, with sudden release $F_e=0$, solved by (\ref{sol_fe}). For $F_e=0$, exact solutions of the solid fraction $\Phi$, fluid velocity $W_l$, solid velocity $W_s$ and internal stress $\Sigma_{zz}$ are calculated from (\ref{phi_sol}, \ref{eq:ws_final}-\ref{Dynamics}). Initial uniform solid fraction of the poroelastic medium is $\Phi_i=0.6$, relaxed solid fraction is $\Phi_r=0.4$, initial height is $H(0)=1$ and $\varepsilon_1^2/\varepsilon_2=0.1$. Part (a) presents change in poroelastic layer height $H-1$ (dash-dotted), external required force $F_e$ (solid), total force by the liquid pressure $F_p$ (dashed) and force applied by the solid surface $F_s$ (dotted). The absolute value of the force applied by the liquid pressure $F_p$ decreases while the sign $F_p$ is negative for all $T$. The force applied by the solid surface $F_s$ decreases in time and is positive for all $T$. The speed $\partial H/\partial T$ is maximal at $T=0$ and decreases with time as the configuration reaches a steady balance. In parts (b-d) blue, purple, red, yellow and green lines mark normalized time $T=0.2$, $0.4$, $0.6$, $0.8$, and $1$, respectively. Part (b) presents solid fraction $\Phi$ vs. $Z$ for various times, showing decrease of $\Phi$ with time and nearly constant values with $Z$. Part (c) presents liquid speed $W_l$ (solid) and solid speed $W_s$ (dashed) vs. $Z$ for various times. Both speeds are approximately linear with $Z$ with gradients decreasing with time. Part (d) presents pressure $P$ (solid) and axial stress $\Sigma_{zz}$ vs. $Z$ for various times. Since no external forces act on the system, the dominant balance is between pressure $P$ and axial stress and thus $P\approx\Sigma_{zz}$. The magnitude of both parameters decreases with time as the system approach steady balance.

\subsubsection{Characteristic values of frogs' toe pads and adhesion time estimation}
We estimate, based on existing works, the order-of-magnitude of the relevant physical properties of frog's toe pads. We focus on torrent frogs due to their ability to keep adhesion underwater and thus without capillary forces. Based on \cite{Federle2006Wet} and \cite{endlein2013stickingunder} we obtain the total toe pad area $10^{-4}m^2$, number of toes $18$, yielding individual toe pad radius of $r_o=1.3\cdot10^{-3}m$ (assuming circular toe pads), frog mass is $2.7\cdot10^{-3}kg$, the viscosity of the mucus is $\mu\approx10^{-3} Pa\cdot s$, toe pad height is $h_r=10^{-5}m$, the permeability of the frog's toe pad is calculated via $k_0\approx d_p^2/4$, where $d_p=10^{-7} m$ is the characteristic length scale of the pores. We examine various values of the characteristic gap in the lubrication region $\tilde{h}$ and the stiffness parameter $m$. For $\tilde{h}=10^{-6}m$, we obtain $\varepsilon_1=h^*/r_o\approx 6.3\cdot10^{-3}$, we  and $\varepsilon_2=k_0 h^*/12\tilde{h}^{3}=0.1$, thus $\varepsilon_1^2/\varepsilon_2\approx 4\cdot10^{-4}$.

Slip will occur when $f_t\geq f_s\mu_s$, where $f_s$ is the dimensional normal force acting on the surface, $\mu_s$ is the friction coefficient and $f_t$ is a tangential force (see Fig. \ref{Figure_1}). Utilizing $F_s=\pi(\Phi_r-\Phi(\Psi=0))$, we can obtain $\Phi_s$, the solid fraction at $\Psi=0$ for which slip will occur
\begin{equation}\label{phi_slip}
\Phi_s=\Phi_r-\frac{1}{\mu_s}\frac{f_t}{m \pi r_o^2}.
\end{equation}
Substituting (\ref{phi_sol}) into (\ref{phi_slip}) yields $H_s$, the height of poroelastic medium when slip occurs
\begin{equation}\label{HS_Pore}
H_s=\frac{\Phi_i}{\Phi_s},
\end{equation}
and substituting (\ref{HS_Pore}) into (\ref{sol_fe}) yields the adhesion time $t_a$. 

\begin{figure}
\includegraphics[width=0.8\textwidth]{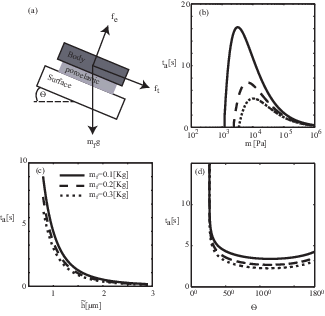}
\centering
\caption{Adhesion time (time until tangential slip occur) $t_a$ for various physical parameters, Part (a) presents the slope angle $\theta$ and the normal $f_e$ and tangential $f_t$ forces. Part (b) presents adhesion $t_a$ time vs. the stiffness of the poroelastic material $m$ for $\theta=60^0$ and $\Phi_i=0.85$. Part (c) presents adhesion $t_a$ time vs. the initial solid fraction $\Phi_i$ (initial compression) for $\theta=60^0$ and $m=5\cdot10^5[Pa]$. Part (d) presents adhesion $t_a$ time vs. slope angle $\theta$ for $\Phi_i=0.85$ and $m=5\cdot10^5[Pa]$. Frog mass of $10^{-3}Kg$ (dashed) $2\cdot10^{-3}Kg$ (dashed) and $3\cdot10^{-3}Kg$ (dotted) are presented. In all cases $\Phi_r=0.7$.}
\label{Figure_5}
\end{figure}

Fig. \ref{Figure_5} presents dimensional adhesion time $t_a$ for various configurations and parameters. Part (a) presents the slope angle $\theta$ and the normal forces. The normal force is 
$f_e=-\cos(\theta)m_f g /n_t$ and the tangential force is $f_t=|\sin(\theta)m_f g /n_t|$, where $m_f$ is the mass of the frog, $g$ is gravity and $n_t$ is the number of frog toes. Parts b-c present adhesion time $t_a$ for different frog mass of $10^{-3}Kg$ (dashed), $2\cdot10^{-3}Kg$ (dashed) and $3\cdot10^{-3}Kg$ (dotted) at $\theta=60^0$. Part b presents adhesion time $t_a$ vs. stiffness parameter $m$ for $\Phi_i=0.85$ and $\theta=60^0$. For stiffness parameter under a critical value ($m\approx10^3Pa$) no adhesion is created. With increasing $m$ a maxima of adhesion time is presented as well as a monotonic decrease in adhesion time as $m\rightarrow\infty$. Part (c) presents adhesion time $t_a$ vs. characteristic gap in the lubrication region $\tilde{h}$ (defined by (\ref{hdif})), for $\theta=60^0$ and $m=5\cdot10^5Pa$. A monotonic increase in adhesion time with decreasing $\tilde{h}$ is observed. Part (d) presents adhesion time vs. slope angle $\theta$ for $\Phi_i=0.85$ and $m=5\cdot10^5 Pa$. Under a critical minimal value of $\theta$ steady adhesion is achieved. With increasing $\theta$ there is a decrease in adhesion time until a minimum value and then a moderate maxima at $\theta=180^0$. In all cases we observe a monotonic increase in adhesion time with reduction of frog mass. 

\section{Concluding Remarks}
In this work we suggested viscous-elastic interaction as a mechanism to explain adhesion of frogs' toe pads. We applied a poroelastic model for the toe pad and a lubrication approximation for the flow between the toe pad and the solid surface. We obtained a governing equation for the solid fraction, which can be solved and then applied to obtain the solid stress, solid velocity, liquid pressure and liquid velocity. The viscous-elastic interaction was shown to be able to create temporary adhesion, even in the absence of capillary forces, and the adhesion time was estimated for physical parameters chosen from order-of-magnitude estimation of frogs' toe pads. A maxima for adhesion time was observed with regard to $m$, the stiffness parameter of the poroelastic medium, and a minima with regard to the slope angle of the surface. The time-dependent nature of viscous-elastic adhesion is in agreement with the periodic repositioning of frogs' toe pads during adhesion to surfaces. This analysis focused on a single toe pad, however, the dynamics of a complete frog includes multiple toe pads which are repositioned at different times. Thus the computed adhesion time $t_a$ represents the time a single toe pad is connected to a surface may differ significantly from adhesion time of the entire frog.

\acknowledgments{This research was supported by the ISRAEL SCIENCE FOUNDATION (Grant 818/13).}

\appendix
\section{Dimensional equations and boundary conditions}
The dimensional governing equation for $\Phi$, the solid fraction, is
\begin{equation}
\frac{\partial \Phi}{\partial t}+\frac{\partial}{\partial z}\left[\Phi\left(\frac{\partial h}{\partial t}-\frac{m k_0}{\mu}\frac{\partial \Phi}{\partial z}\right)\right]=0
\end{equation} 
with the dimensional boundary conditions
\begin{equation}
\frac{\partial \Phi}{\partial z}-\frac{\mu}{m k_0}\frac{\partial h}{\partial t}\Phi=0, \,\,\, z=0
\end{equation}
\begin{equation}
\frac{\partial \Phi}{\partial z}=0,\,\,\, z=h(t)
\end{equation}
and initial condition
\begin{equation}
\Phi=\Phi_i,\,\,\, t=0,
\end{equation}
where the parameters $k_0$, $m$ and $\Phi_r$ represent permeability, stiffness parameter and relaxed solid fraction, respectively, are known constants defining the properties of the poroelastic medium by the relations \citep[following][]{Anderson2005Imbibition,siddique2009newtonian,Siddique2009Capillary}
\begin{equation}\label{eq:streconst_Ap}
k_{zz}=\frac{k_0}{\Phi},\,\,\, \sigma=m(\Phi_r-\Phi).
\end{equation}

The dimensional liquid speed $w_l$ and solid speed $w_s$ are
\begin{equation}
w_s=\frac{\partial h}{\partial t}-\frac{m k_0}{\mu(1-\Phi)}\frac{\partial \Phi}{\partial z}
\end{equation} 
and
\begin{equation}
w_l=\frac{\partial h}{\partial t}+\frac{m k_0}{\mu\Phi}\frac{\partial \Phi}{\partial z}.
\end{equation} 
The dimensional gauge pressure at $z=0$ is
\begin{equation}
p(z=0)=-3\frac{\mu r_o^2}{\tilde{h}^{3}}\left(1-\Phi(0)\right)\left(\frac{\partial h}{\partial t}+\frac{m k_0}{\mu (1-\Phi(0))}\frac{\partial \Phi(0)}{\partial z}\right)\left(1-\frac{r^2}{r_o^2}\right).
\end{equation} 
The dimensional force balance equation of the poroelastic medium is
\begin{equation}
f_e=\frac{3\pi}{2}\frac{\mu r_o^2}{\tilde{h}^{3}}\left(1-\Phi(0)\right)\left(\frac{\partial h}{\partial t}+\frac{m k_0}{\mu (1-\Phi(0))}\frac{\partial \Phi(0)}{\partial z}\right)-m\pi r_o^2(\Phi(z=0)-\Phi_r).
\end{equation} 
where $f_e$ is the external force. Slip occurs when $f_t>\mu_s f_s$, where $f_s = m \pi r_o^2 ( \Phi (z=0) - \Phi _r) $ is the normal force applied by the solid fraction of the poroelastic material and and $\mu_s$ is the friction coefficient.

For $\varepsilon_1^2/\varepsilon_2\rightarrow 0$ we obtain the relation between $\Phi$ and $h$
\begin{equation}
\frac{\Phi}{\Phi_i}\sim\frac{h(0)}{h}+\frac{\mu h(0)}{m k_0}\frac{\partial h}{\partial t}\left(\frac{z}{h}-\frac{1}{2}\frac{z^2}{h^2}-\frac{1}{3}\right).
\end{equation}

\bibliographystyle{jfm}
\bibliography{Bib_File}

\begin{thebibliography}{25}
\expandafter\ifx\csname natexlab\endcsname\relax\def\natexlab#1{#1}\fi

\bibitem[Ambrosi \& Preziosi(2000)]{Ambrosi2000Modeling}
{\sc Ambrosi, D. \& Preziosi, L.} 2000 Modeling injection molding processes
  with deformable porous preforms. {\em SIAM J. Appl. Math.\/} {\bf 61}~(1),
  22--42.

\bibitem[Anderson(2005)]{Anderson2005Imbibition}
{\sc Anderson, D.M.} 2005 Imbibition of a liquid droplet on a deformable porous
  substrate. {\em Physics of Fluids\/} {\bf 17}~(8), 087104--087104--22.

\bibitem[Atkin \& Craine(1976)]{atkin1976continuum}
{\sc Atkin, R.J. \& Craine, R.E.} 1976 Continuum theories of mixtures: basic
  theory and historical development. {\em Quarterly J. Mech. and Appl. Math.\/}
  {\bf 29}~(2), 209--244.

\bibitem[Barnes {\em et~al.\/}(2002)Barnes, Smith, Oines \&
  Mundl]{barnes2002bionics}
{\sc Barnes, J, Smith, J, Oines, C \& Mundl, R} 2002 Bionics and wet grip. {\em
  Tire Technology International\/} {\bf 2002}~(Dec), 56--60.

\bibitem[Battiato(2012)]{battiato2012self}
{\sc Battiato, Ilenia} 2012 Self-similarity in coupled brinkman/navier--stokes
  flows. {\em Journal of Fluid Mechanics\/} {\bf 699}, 94--114.

\bibitem[Battiato {\em et~al.\/}(2010)Battiato, Bandaru \&
  Tartakovsky]{battiato2010elastic}
{\sc Battiato, Ilenia, Bandaru, Prabhakar~R \& Tartakovsky, Daniel~M} 2010
  Elastic response of carbon nanotube forests to aerodynamic stresses. {\em
  Physical review letters\/} {\bf 105}~(14), 144504.

\bibitem[Beavers \& Joseph(1967)]{Beavers.1967}
{\sc Beavers, G.S. \& Joseph, D.D.} 1967 {Boundary conditions at a naturally
  permeable wall}. {\em J. Fluid Mech.\/} {\bf 30}, 197--207.

\bibitem[Biot(1972)]{Biot1972Theory}
{\sc Biot, M.A.} 1972 Theory of finite deformations of porous solids. {\em
  Indiana University Math. J.\/} {\bf 21}~(7), 597--620.

\bibitem[Bowen(1980)]{Bowen1980Incompressible}
{\sc Bowen, R.M.} 1980 Incompressible porous media models by use of the theory
  of mixtures. {\em Int. J. Eng. Sci.\/} {\bf 18}~(9), 1129--1148.

\bibitem[Emerson \& Diehl(1980)]{Emerson1980Toe}
{\sc Emerson, S.B. \& Diehl, D.} 1980 Toe pad morphology and mechanisms of
  sticking in frogs. {\em Biological J. Linnean Society\/} {\bf 13}~(3),
  199--216.

\bibitem[Endlein {\em et~al.\/}(2013{\natexlab{{\em a\/}}})Endlein, Barnes,
  Samuel, Crawford, Biaw \& Grafe]{endlein2013stickingunder}
{\sc Endlein, T., Barnes, W. J.~P., Samuel, D.~S., Crawford, N.~A., Biaw, A.~B.
  \& Grafe, U.} 2013{\natexlab{{\em a\/}}} Sticking under wet conditions: The
  remarkable attachment abilities of the torrent frog, staurois guttatus. {\em
  PLOS ONE\/} {\bf 8}~(9), e73810.

\bibitem[Endlein {\em et~al.\/}(2013{\natexlab{{\em b\/}}})Endlein, Ji, Samuel,
  Yao, Wang, Barnes, Federle, Kappl \& Dai]{endlein2013stickinglike}
{\sc Endlein, T., Ji, A., Samuel, D., Yao, N., Wang, Z., Barnes, W. J.~P.,
  Federle, W., Kappl, M. \& Dai, Z.} 2013{\natexlab{{\em b\/}}} Sticking like
  sticky tape: tree frogs use friction forces to enhance attachment on
  overhanging surfaces. {\em J. Royal Society Interface\/} {\bf 10}~(80),
  20120838.

\bibitem[Ernst(1973{\natexlab{{\em a\/}}})]{ernst1973digitalI}
{\sc Ernst, V.V.} 1973{\natexlab{{\em a\/}}} The digital pads of the tree frog,
  hyla cinerea. i. the epidermis. {\em Tissue and Cell\/} {\bf 5}~(1), 83--96.

\bibitem[Ernst(1973{\natexlab{{\em b\/}}})]{ernst1973digitalII}
{\sc Ernst, V.V} 1973{\natexlab{{\em b\/}}} The digital pads of the tree frog,
  hyla cinerea. ii. the mucous glands. {\em Tissue and Cell\/} {\bf 5}~(1),
  97--104.

\bibitem[Federle(2006)]{Federle2006Why}
{\sc Federle, W.} 2006 Why are so many adhesive pads hairy? {\em J. Exp.
  Biology\/} {\bf 209}~(Pt 14), 2611--2621, lR: 20131121; JID: 0243705;
  ppublish.

\bibitem[Federle {\em et~al.\/}(2006)Federle, Barnes, Baumgartner, Drechsler \&
  Smith]{Federle2006Wet}
{\sc Federle, W., Barnes, W.J., Baumgartner, W., Drechsler, P. \& Smith, J.M.}
  2006 Wet but not slippery: Boundary friction in tree frog adhesive toe pads.
  {\em J. Royal Society, Interface\/} {\bf 3}~(10), 689--697, lR: 20130904; GR:
  Wellcome Trust/United Kingdom; JID: 101217269; OID: NLM: PMC1664653;
  ppublish.

\bibitem[Gat {\em et~al.\/}(2011)Gat, Navaz \& Gharib]{Gat.2011}
{\sc Gat, A.D., Navaz, H. \& Gharib, M.} 2011 Dynamics of freely moving plates
  connected by a shallow liquid bridge. {\em Physics of Fluids
  (1994-present)\/} {\bf 23}~(9), 097101.

\bibitem[Green(1979)]{green1979treefrog}
{\sc Green, D.M.} 1979 Treefrog toe pads: comparative surface morphology using
  scanning electron microscopy. {\em Canadian J. Zoology\/} {\bf 57}~(10),
  2033--2046.

\bibitem[Hanna {\em et~al.\/}(1991)Hanna, Jon \& Barnes]{HANNA1991Adhesion}
{\sc Hanna, G., Jon, W. \& Barnes, W.J.} 1991 Adhesion and detachment of the
  toe pads of tree frogs. {\em J. Exp. Biology\/} {\bf 155}~(1), 103--125.

\bibitem[Persson(2007)]{persson2007wet}
{\sc Persson, BNJ} 2007 Wet adhesion with application to tree frog adhesive toe
  pads and tires. {\em Journal of Physics: Condensed Matter\/} {\bf 19}~(37),
  376110.

\bibitem[Preziosi {\em et~al.\/}(1996)Preziosi, Joseph \&
  Beavers]{Preziosi1996Infiltration}
{\sc Preziosi, L., Joseph, D.D. \& Beavers, G.S.} 1996 Infiltration of
  initially dry, deformable porous media. {\em Int. J. Multiphase Flow\/} {\bf
  22}~(6), 1205--1222.

\bibitem[Rajagopal \& Tao(1995)]{rajagopalmechanics}
{\sc Rajagopal, K.R. \& Tao, L.} 1995 Mechanics of mixtures. {\em World Sci.,
  Singapore\/} .

\bibitem[Siddique(2009)]{siddique2009newtonian}
{\sc Siddique, J.I.} 2009 Newtonian and non-newtonian flows into deformable
  porous materials. PhD thesis, George Mason University.

\bibitem[Siddique {\em et~al.\/}(2009)Siddique, Anderson \&
  Bondarev]{Siddique2009Capillary}
{\sc Siddique, J.I., Anderson, D.M. \& Bondarev, A.} 2009 Capillary rise of a
  liquid into a deformable porous material. {\em Physics of Fluids\/} {\bf
  21}~(1), 013106.

\bibitem[Tsipenyuk \& Varenberg(2014)]{tsipenyuk2014use}
{\sc Tsipenyuk, A. \& Varenberg, M.} 2014 Use of biomimetic hexagonal surface
  texture in friction against lubricated skin. {\em J. Royal Soc. Interface\/}
  {\bf 11}, 20140113.

\end{thebibliography}

\end{document}